\documentclass[review,3p,times]{elsarticle_accepted}

\usepackage{amssymb}
\usepackage{pdflscape}
\usepackage{amsmath}
\usepackage{amsthm}
\usepackage{empheq}

\usepackage[nodots]{numcompress}
\usepackage{nomencl}
\usepackage{framed} 
\usepackage{multicol} 
\makenomenclature

\usepackage{lineno}
\usepackage[english]{babel}

\journal{Ultrasonics}

\begin{document}

\begin{frontmatter}



\title{Wave steering effects in anisotropic composite structures: Direct calculation of the energy skew angle through a finite element scheme}


\author[Nottingham]{D. Chronopoulos}
\ead{Dimitrios.Chronopoulos@nottingham.ac.uk}

\address[Nottingham]{Institute for Aerospace Technology $\&$ The Composites Group, The University of Nottingham,  NG7 2RD, UK}

\begin{abstract}
A systematic expression quantifying the wave energy skewing phenomenon as a function of the mechanical characteristics of a non-isotropic structure is derived in this study. A structure of arbitrary anisotropy, layering and geometric complexity is modelled through Finite Elements (FEs) coupled to a periodic structure wave scheme. A generic approach for efficiently computing the angular sensitivity of the wave slowness for each wave type, direction and frequency is presented. The approach does not involve any finite differentiation scheme and is therefore computationally efficient and not prone to the associated numerical errors.

\end{abstract}

\begin{keyword}
Wave Steering \sep Composite Structures \sep Energy skewing \sep Caustics \sep Wave Finite Elements


\end{keyword}

\end{frontmatter}



\begin{table*}
  \begin{framed}
\begin{tabular}{l l}
\textbf{Nomenclature} \\ \\
${\bf B}$ & Shape function derivative matrix of a single FE \\
$\mathbf{C}_0$ & Elastic stiffness matrix at the material principal axis\\
${\bf J}$& Jacobian matrix of a single FE \\
$\mathcal{K}$ & Intermediate stiffness matrix employed for the assembly of $\mathbb{K}$\\
$\mathbb{M}$, $\mathbb{K}$ & Mass and stiffness matrices of the periodic element \\
${\bf R}$ & Displacement phase transformation matrix \\
$\mathbf{T}$ & Coordinate transformation matrix \\
${\bf k}$ & Stiffness matrix of a single FE \\
${\bf q}$ & Physical displacement vector for the elastic waveguide \\
$L_x$, $L_y$ & Dimensions of the modelled periodic segment \\
$L$, $R$, $B$, $T$, $I$ & Left, right, bottom, top sides and interior indices \\
$N$ & Number of elements \\
$c_g$ & Group velocity \\
$k$ & Wavenumber\\
$l_x, l_y, l_z$ & Dimensions of a single FE \\
$s$ & Wave slowness\\
$w$ & Wave type index\\
\\
$\mathbf{x}$ & Wave mode shape vector for the elastic waveguide \\
$\varepsilon$ & Propagation constant \\
$\theta$ & Wave propagation angle \\
$\eta$, $\xi$, $\mu$ & Local FE coordinates \\
$\lambda$ & Eigenvalue of the wave propagation eigenproblem\\
$\psi$ & Energy skew angle\\
$\xi$ & Coordinate transformation angle \\
$\omega$ & Angular frequency\\
\end{tabular}
  \end{framed}
\end{table*}

\section{Introduction}
\label{sec:oduct}

Understanding complex wave phenomena is of paramount importance for the successful application of ultrasonic techniques within the non-destructive testing (NDT) and biomedical fields. Accurate and efficient modelling of elastic wave propagation complex phenomena in composite structures play a crucial role in the development of robust algorithms for damage detection and localization. One of the most prominent of these phenomena is the so-called energy skewing (see Fig.\ref{fig:fig1.eps}), induced by the angular divergence between the phase and group velocities for non-isotropic configurations. Wave skewing results in a non-uniform distribution of energy along the wavefront. An inaccurate description of the skewing effect in the computational models and NDT algorithms can well result in an incorrect prediction of damage location \cite{yan2010ultrasonic,kersemans2014pitfalls} and type.

Directional dependence of the wave slowness characteristics in non-isotropic structures has been well discussed and investigated by several researchers. In \cite{lean1979large} the authors demonstrated a material anisotropy-based, beam-steering scheme for electronically steering an acoustic beam over an angle larger than 70$^o$ in a TeO$_2$ crystal. The idea was based on the pronounced angular dependency of the wave skewing angle in the same material. Wave beam steering through the employment of phased array transducers \cite{turnbull1991beam} has been discussed within the context of several applications including biomedical imaging \cite{smith1991high}, structural health monitoring \cite{clay1999experimental,wooh1999optimum,salas2009guided} and acoustic applications \cite{wu2012fpga}. With regard to layered cellular composites, the researchers in \cite{ruzzene2003wave,hussein2008wave,casadei2013anisotropy} derived wave propagation models based on Bloch's theorem in order to show how band-gaps and strong acoustic focusing can be affected by structural anisotropy in periodic lattice structures.

Calculation of the wavefront curve has formed the basis for most researchers in order to quantify wave steering effects. The wave skewing angle has been calculated by a number of authors through a variety of approaches, including the application of a Fresnel approximation to the wave propagation problem \cite{newberry1989paraxial}, derivation through the propagating group velocities in two orthogonal directions within the panel \cite{rose2004ultrasonic}, as well as through a Finite Differentiation (FD) approach \cite{wang2007group}. To the best of the author's knowledge, there is currently no expression directly quantifying the wave skewing effect as a function of the mechanical characteristics of the non-isotropic structure.

The principal objective and contributing novelty of this study is the derivation of a systematic and robust expression relating the wave energy skew angle to the material characteristics of the composite structure under investigation. A robust FE-based approach for efficiently computing the angular sensitivity of the wave phase velocities for each wave type, direction and frequency is presented. The considered structure can be of arbitrary layering and material characteristics as FE modelling is employed. The exhibited scheme is able to compute the wavenumber angular sensitivity (and subsequently the energy skew angle) by determining and post-processing a single solution of the system. This overcomes the drawbacks of the currently employed FD approaches.

The paper is organized as follows: In Sec.\ref{sec:psicomputsoi} a general expression is derived for the angle of the propagating energy wavefront as well as the skew angle between the phase and group velocities for each wave type as a function of the wavenumber angular sensitivity. In Sec.\ref{sec:avesensitivitya} a direct expression of the wavenumber sensitivity with respect to the direction of propagation is derived within a FE modelling context.  Numerical case studies validating the computational scheme are presented in Sec.\ref{sec:numericcasstudn}. Conclusions on the exhibited work are eventually drawn in Sec.\ref{sec:Conclusions}.

\section{Calculation of the wave energy skew angle}
\label{sec:psicomputsoi}

\begin{figure}
  \centering
  \includegraphics[clip,width=10cm]{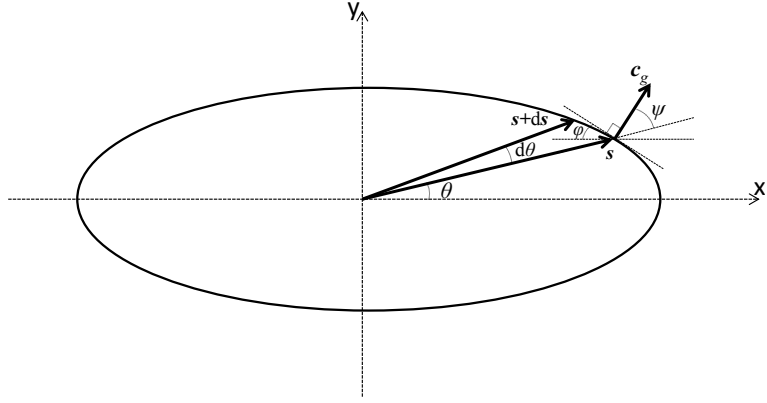}
  \caption{Illustration of the group velocity being perpendicular to the wave slowness curve for a non-isotropic structure. A wave energy skew angle $\psi$ is thus formed. An infinitesimal change of angle $\textnormal{d} \theta$ and slowness $\textnormal{d} s$ is also shown. The angle $\phi$ is formed between the horizontal and the tangent.}
\label{fig:fig1.eps}
\end{figure}

Slowness curves are particularly useful for visualizing the direction of the group velocity (see Fig.\ref{fig:fig1.eps}). On the other hand, the velocity of the wavefront (defined as the locus of ray velocity vectors along all directions starting from the origin) in the direction normal to the wavefront is known as the phase velocity. In an anisotropic material, the phase and group velocities are generally different \cite{carcione2007wave} and a clear distinction between the two should be made to ensure that the correct velocity profile is employed when performing health monitoring with an ultrasonic device. The physical difference between the phase and group velocities can be described by considering a propagating wave packet (see Fig.\ref{fig:fig1.eps}). The wavefronts remain normal to the the phase velocity direction $\theta$ (or equivalently, parallel to the transducer surface exciting the packet), however due to material anisotropy the wave packet skews away from the normal direction by an angle $\psi$ and instead travels along a shifted ray path. The velocity of the wave packet envelope is given by the group velocity $c_{g}$. It has been well documented \cite{rose2004ultrasonic} that the group velocity vector is always perpendicular to the tangent of the slowness curve. Moreover, it is reminded that the slowness of a wave $w$ can be expressed as $\displaystyle s_w=\frac{k_w}{\omega_w}$.

When the angular rate of change for each propagating wavenumber $k_w$ is known (see Sec.\ref{sec:avesensitivitya}), the skew angle $\psi_w$ can be determined through geometric considerations. In Fig.\ref{fig:fig1.eps}, a representation of an infinitesimal change of angle $\textnormal{d} \theta$ and correspondingly of slowness $\textnormal{d} s_w$ is drawn. In the same figure the angle of the tangent to the slowness with respect to the horizontal $\phi$ is shown. As vector $c_g$ is perpendicular to the drawn tangent and $s_w$ forms an angle $\theta$ to the horizontal, the skew angle $\psi_w$ can be determined as
\begin{subequations}
\begin{empheq}{align}
 \psi_w&=\frac{\pi}{2}-\theta-\phi_w  \quad \text{ $0\leq\theta<\pi$ } \\
 \psi_w&=\frac{3\pi}{2}-\theta-\phi_w \quad \text{ $\pi\leq \theta<2\pi$ }
\end{empheq}
\label{systemofmodalavererqusd33}
\end{subequations}
\noindent
It is straightforward to deduce that
\begin{equation}
\tan(\phi)=\frac{(s+ \textnormal{d} s) \sin(\theta+\textnormal{d} \theta)-s \sin\theta}{s \cos \theta-(s+\textnormal{d} s) \cos(\theta+\textnormal{d} \theta)}=\frac{(k+ \textnormal{d} k) \sin(\theta+\textnormal{d} \theta)-k \sin\theta}{k \cos \theta-(k+\textnormal{d} k) \cos(\theta+\textnormal{d} \theta)}
\end{equation}
\noindent
which after expanding the sine and cosine terms using the appropriate identities and employing infinitesimal angles approximations can be written as
\begin{equation}
\tan(\phi)=\frac{(k+ \textnormal{d} k) (\sin\theta+\cos(\theta)\textnormal{d}\theta)-k \sin\theta}{k \cos \theta-(k+\textnormal{d} k) (\cos\theta-\sin(\theta)\textnormal{d}\theta)}
\end{equation}
\noindent
Dividing the above expression by $\cos(\theta)\textnormal{d} \theta$, eventually gives
\begin{equation}
\tan(\phi)=\frac{\tan\theta \displaystyle \frac{\textnormal{d} k}{\textnormal{d}\theta}+k} {k\tan\theta-\displaystyle \frac{\textnormal{d} k}{\textnormal{d}\theta}}
\label{notewmneewr}
\end{equation}
\noindent
A number of numerical and analytical techniques can be used to compute the directional wavenumbers $k(\theta)$ (see \ref{sec:vityanal} for the one used in this work). The following section provides a concise expression for the angular wavenumber sensitivity expression $\displaystyle \frac{\textnormal{d} k}{\textnormal{d}\theta}$.

\section{Angular sensitivity of the wave phase velocity in an anisotropic composite}
\label{sec:avesensitivitya}

\begin{figure}
  \centering
  \includegraphics[clip,width=5.5cm]{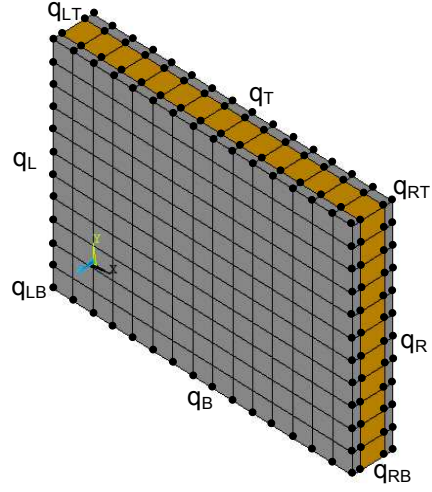}
  \caption{Caption of a FE modelled composite layered panel}
\label{fig:stif_panel.eps}
\end{figure}

A periodic segment of a composite panel having arbitrary layering and material characteristics is hereby considered (see Fig.\ref{fig:stif_panel.eps}) with $L_x$, $L_y$ its dimensions in the $x$ and $y$ directions respectively. The structural segment can be modelled using a conventional FE package and the mass and stiffness matrices of the segment $\mathbb{M}$, $\mathbb{K}$ can be computed in a straightforward manner. A periodic structure wave scheme can be employed in order to numerically determine the propagating wavenumbers $k_w$ and the corresponding mode shapes $\mathbf{x}_w$ for each propagating wave mode type as exhibited in \ref{sec:vityanal}.

It is noted that matrices $\mathbf{K}={\bf R}^*\mathbb{K}{\bf R}$ and $\mathbf{M}={\bf R}^*\mathbb{M}{\bf R}$ in Eq.(\ref{eq:eigen}) are Hermitian therefore their resulting eigenvalues are real and the set of eigenvectors will be orthogonal. Eigenvalue sensitivity for standard eigenproblems is an established result in modern literature \cite{nelson1976simplified,adhikari2001eigenderivative} that will be employed in the present work. The eigenproblem in Eq.\ref{eq:eigen} can be differentiated with respect to the angle of wave propagation $\theta$ giving
\begin{equation}
[\mathbf{ K}  -  \lambda_w \mathbf{ M}] \frac{ \partial \mathbf{x}_{w}}{\partial \theta} + \left( \frac{ \mathbf{\partial K}}{\partial \theta} - \lambda_{w}\frac{ \mathbf{\partial M}}{\partial \theta} \right)\mathbf{x}_{w} - \frac{\partial \lambda_w}{\partial \theta} \mathbf{ M}\mathbf{x}_{w}=\mathbf{ 0}
\label{eq:almodfhiuwmosdieord58}
\end{equation}
\noindent
After multiplying the above expression by $\mathbf{x}^\top_{w}$ and making use of the mass normalization of the eigenmodes the following expression can be derived for the angular sensitivity of the computed eigenalues
\begin{equation}
\frac{\partial \lambda_w}{\partial \theta}  = \mathbf{x}^\top_{w}      \left( \frac{ \mathbf{\partial K}}{\partial \theta} - \lambda_{w}\frac{ \mathbf{\partial M}}{\partial \theta} \right)\mathbf{x}_{w}
\label{eq:almodfhiuw}
\end{equation}
In case of repeated eigenvalues being detected, the sensitivity expression should be modified according to the findings in \cite{juang1989eigenvalue,friswell1996derivatives}. Taking into account that $\mathbb{M}$ and $\mathbb{K}$ have no angular dependence, the above expression can be developed to provide a more generic angular eigenvalue sensitivity expression
\begin{equation}
\frac{\partial \lambda_w}{\partial \theta}  = \mathbf{x}^\top_{w}      \frac{ \mathbf{\partial K}}{\partial \theta} \mathbf{x}_{w} - \lambda_{w}\mathbf{x}^\top_{w}      \frac{ \mathbf{\partial M}}{\partial \theta} \mathbf{x}_{w} = \mathbf{x}^\top_{w} \left( \displaystyle \frac{\partial {\bf R}^*}{\partial \theta}\mathbb{K}{\bf R}+   {\bf R}^*\mathbb{K}\displaystyle \frac{\partial {\bf R}}{\partial \theta}\right)\mathbf{x}_{w} - \lambda_{w}\mathbf{x}^\top_{w} \left( \displaystyle \frac{\partial {\bf R}^*}{\partial \theta}\mathbb{M}{\bf R}+{\bf R}^*\mathbb{M}\displaystyle \frac{\partial {\bf R}}{\partial \theta} \right) \mathbf{x}_{w}
\label{eq:almodfhiuwmosdieord}
\end{equation}
\noindent
For the wavenumber sensitivity $\displaystyle\frac{\partial k_w}{\partial \theta}$ the following expression stands
\begin{equation}
\frac{\partial k_w}{\partial \theta} =\frac{\partial k_w}{\partial \omega_w} \frac{\partial \omega_w}{\partial \lambda_w} \frac{\partial \lambda_w}{\partial \theta}
\label{angsenwanbe}
\end{equation}
\noindent
while the inverse of the group velocity $\displaystyle\frac{\partial k_w}{\partial \omega_w}$ can be computed \cite{finnveden2004evaluation,Cotoni,ichchou2007guided} directly through the results of a single eigenvalue solution (that is avoiding FD for one more time) by differentiating the eigenproblem in Eq.\ref{eq:eigen} with respect to $k_w$, deriving
\begin{equation}
\left( \displaystyle \frac{\partial \mathbf{R}^{*}}{\partial k_w} [\mathbb{K}-\omega_w^2\mathbb{M}] {\bf R} +  {\bf R}^* [\mathbb{K}-\omega_w^2\mathbb{M}] \displaystyle \frac{\partial \mathbf{R}}{\partial k_w} -  2 \omega_w \frac{\partial \omega_w}{\partial k_w} \mathbf{R}^{*} \mathbb{M}  {\bf R} \right) \mathbf{x}_{w} + \mathbf{R}^{*} [\mathbb{K}-\omega_w^2\mathbb{M}] {\bf R} \frac{\partial \mathbf{x}_{w}}{\partial k_w}   = {\bf 0}
\end{equation}
\noindent
and by multiplying the above expression by $\mathbf{x}^\top_{w}$ and taking advantage of the orthogonality properties the $\displaystyle\frac{\partial k_w}{\partial \omega_w}$ term can be directly obtained as
\begin{equation}
\frac{\partial k_w}{\partial \omega_w}  =       \left( \displaystyle \frac{ 2 \omega_w}{\mathbf{x}^\top_{w} \left( \displaystyle \frac{\partial \mathbf{R}^{*}}{\partial k_w} [\mathbb{K}-\omega_w^2\mathbb{M}] {\bf R}     +       {\bf R}^* [\mathbb{K}-\omega_w^2\mathbb{M}] \displaystyle \frac{\partial \mathbf{R}}{\partial k_w}            \right) \mathbf{x}_{w}} \right)
\label{invgroupvel}
\end{equation}
\noindent
Eventually (taking into account that $\displaystyle\frac{\partial \omega_w}{\partial \lambda_w}  = \frac{1}{2\omega_w}$),  Eq.\ref{angsenwanbe} can therefore provide a direct expression of the angular wavenumber sensitivity for any propagating wave type $w$ and direction of propagation $\theta$ at angular frequency $\omega_w$
\begin{equation}
\displaystyle \frac{\partial k_w}{\partial \theta}  =       \left( \displaystyle \frac{  \mathbf{x}^\top_{w} \left( \displaystyle \frac{\partial {\bf R}^*}{\partial \theta}\mathbb{K}{\bf R}+{\bf R}^*\mathbb{K}\displaystyle \frac{\partial {\bf R}}{\partial \theta}\right)\mathbf{x}_{w} - \lambda_{w}\mathbf{x}^\top_{w} \left( \displaystyle \frac{\partial {\bf R}^*}{\partial \theta}\mathbb{M}{\bf R}+{\bf R}^*\mathbb{M}\displaystyle \frac{\partial {\bf R}}{\partial \theta} \right) \mathbf{x}_{w} }{\mathbf{x}^\top_{w} \left( \displaystyle \frac{\partial \mathbf{R}^{*}}{\partial k_w} [\mathbb{K}-\omega_w^2\mathbb{M}] {\bf R}     +       {\bf R}^* [\mathbb{K}-\omega_w^2\mathbb{M}] \displaystyle \frac{\partial \mathbf{R}}{\partial k_w}            \right) \mathbf{x}_{w}} \right)
\label{4uthng}
\end{equation}
It is noted that ${\bf R}$ is a direct function of $k_w$ and $\theta$, therefore the $\displaystyle \frac{\partial \mathbf{R}}{\partial k_w}$ and $\displaystyle \frac{\partial \mathbf{R}}{\partial \theta}$ terms are straightforward \cite{chronopoulos2015design,chronopoulos2014predicting,chronopoulos2017wave} to compute. The global stiffness matrix $\mathbb{K}$ of the structural segment is formed by adding the local stiffness matrices of individual FEs as
\begin{equation}
\mathbb{K}=\sum_{p=1}^N\mathcal{K}_p \ \ \ \text{with}  \ \ \  \mathcal{K}^{[(3p-2):3p,(3p-2):3p]}_p = {\bf k}_p
\end{equation}
\noindent
with $N$ the total number of FEs and the superscript of $\mathcal{K}_p$ denoting the exact positioning of ${\bf k}_p$ within it. The remaining entries in $\mathcal{K}_p$ are null. The individual FE stiffness matrices can be computed as
\begin{equation}
{\bf k}_p=\int_{-1}^1\int_{-1}^1\int_{-1}^1 {\bf B}^{\top}\mathbf{C}_0{\bf B}\vert{\bf J}\vert \ \textrm{d}\eta \textrm{d}\xi \textrm{d}\mu
\end{equation}
\noindent
with ${\bf J}$ the Jacobian and ${\bf B}$ the shape function derivative matrices of the element, while $\mathbf{C}_0$ is the elastic stiffness matrix at the material principal axis which can contain up to 21 independent coefficients (for a triclinic material), input as
\begin{equation}
\mathbf{C}_0=\left[ {\begin{array}{cccccc}
c_{11} &c_{12} &c_{13} &c_{14} &c_{15} &c_{16} \\
c_{12} &c_{22} &c_{23} &c_{24} &c_{25} &c_{26}\\
c_{13} &c_{23} &c_{33} &c_{34} &c_{35} &c_{36} \\
c_{14} &c_{24} &c_{34} &c_{44} &c_{45} &c_{46} \\
c_{15} &c_{25} &c_{35} &c_{45} &c_{55} &c_{56} \\
c_{16} &c_{26} &c_{36} &c_{46} &c_{56} &c_{66} \end{array} } \right]
\end{equation}
\noindent
If a revolution angle $\xi$ is considered between the material principal axis and the effective transformed coordinate system, then the transformed elastic stiffness matrix (rotated about $z$ axis) can be calculated as \cite{jones1975mechanics}
\begin{equation}
\mathbf{C}=\mathbf{T}^{-1} \mathbf{C}_0 \mathbf{T}^{-\top}
\end{equation}
\noindent
with $\mathbf{T}^{-1}$ being the inverse of the coordinate transformation matrix given by
\small
\begin{equation}
\mathbf{T}^{-1}=\left[ {\begin{array}{cccccc}
\cos^2(-\xi) & \sin^2(-\xi) &0 &0 &0 &2\cos(-\xi)\sin(-\xi)\\
    \sin^2(-\xi) & \cos^2(-\xi) &0 &0 &0 &-2\cos(-\xi)\sin(-\xi)\\
    0& 0& 1& 0& 0& 0\\
    0& 0& 0& \cos(-\xi)& -\sin(-\xi)& 0\\
    0& 0& 0& \sin(-\xi)& \cos(-\xi)& 0\\
    -\cos(-\xi)\sin(-\xi) &\cos(-\xi)\sin(-\xi) &0 &0 &0 &\cos^2(-\xi)-\sin^2(-\xi) \end{array} } \right]
\end{equation}
\normalsize
Eventually, substituting Eq.\ref{4uthng} into Eq.\ref{notewmneewr} and subsequently into Eq.\ref{systemofmodalavererqusd33} provides a generic expression of the energy skew angle for each wave type $w$ as
\begin{equation}
\psi_w=\displaystyle \frac{\pi}{2}-\theta-          \arctan \left( \frac{\tan\theta \displaystyle \left( \displaystyle \frac{  \mathbf{x}^\top_{w} \left( \displaystyle \frac{\partial {\bf R}^*}{\partial \theta}\mathbb{K}{\bf R}+   {\bf R}^*\mathbb{K}\displaystyle \frac{\partial {\bf R}}{\partial \theta}\right)\mathbf{x}_{w} - \lambda_{w}\mathbf{x}^\top_{w} \left( \displaystyle \frac{\partial {\bf R}^*}{\partial \theta}\mathbb{M}{\bf R}+{\bf R}^*\mathbb{M}\displaystyle \frac{\partial {\bf R}}{\partial \theta} \right) \mathbf{x}_{w} }{\mathbf{x}^\top_{w} \left( \displaystyle \frac{\partial \mathbf{R}^{*}}{\partial k_w} [\mathbb{K}-\omega_w^2\mathbb{M}] {\bf R}     +       {\bf R}^* [\mathbb{K}-\omega_w^2\mathbb{M}] \displaystyle \frac{\partial \mathbf{R}}{\partial k_w}            \right) \mathbf{x}_{w}}  \right)+k_w} {k_w \tan\theta-\displaystyle \left( \displaystyle \frac{  \mathbf{x}^\top_{w} \left( \displaystyle \frac{\partial {\bf R}^*}{\partial \theta}\mathbb{K}{\bf R}+   {\bf R}^*\mathbb{K}\displaystyle \frac{\partial {\bf R}}{\partial \theta}\right)\mathbf{x}_{w} - \lambda_{w}\mathbf{x}^\top_{w} \left( \displaystyle \frac{\partial {\bf R}^*}{\partial \theta}\mathbb{M}{\bf R}+{\bf R}^*\mathbb{M}\displaystyle \frac{\partial {\bf R}}{\partial \theta} \right) \mathbf{x}_{w} }{\mathbf{x}^\top_{w} \left( \displaystyle \frac{\partial \mathbf{R}^{*}}{\partial k_w} [\mathbb{K}-\omega_w^2\mathbb{M}] {\bf R}     +       {\bf R}^* [\mathbb{K}-\omega_w^2\mathbb{M}] \displaystyle \frac{\partial \mathbf{R}}{\partial k_w}            \right) \mathbf{x}_{w}}  \right)} \right)
\label{finrewmpodsi}
\end{equation}
\noindent
which quantifies the wave energy skewing as a direct function of the mechanical characteristics of the layered structure. It is reminded that the above expression is valid for $0\leq\theta<\pi$ (see Eq.\ref{systemofmodalavererqusd33} for the remaining quadrants).

\section{Numerical case studies}
\label{sec:numericcasstudn}

\begin{figure}
  \centering
\includegraphics[clip,width=13cm]{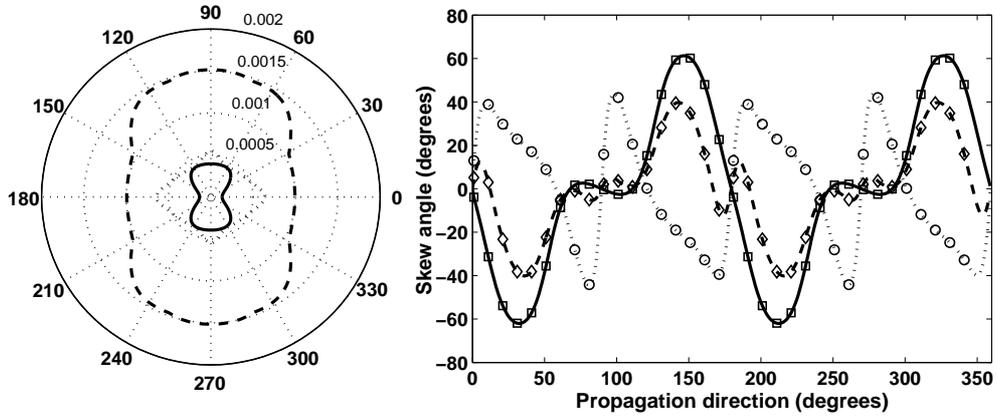}
  \caption{Left: Wave slowness curves for the P$_0$ (--), SH$_0$ ($\cdots$) and A$_0$ (-\,-) waves propagating in the orthotropic graphite-epoxy monolithic structure at 0.1MHz. Right: Corresponding energy skew angles computed through the presented approach for the P$_0$ (--), SH$_0$ ($\cdots$) and A$_0$ (-\,-) waves. Also presented the skew angles computed through a FD scheme as exhibited in \cite{wang2007group} for the P$_0$ ($\square$), SH$_0$ ($\circ$) and A$_0$ ($\diamond$) waves. }
\label{fr1_2.eps}
\end{figure}

\begin{figure}
  \centering
\includegraphics[clip,width=13cm]{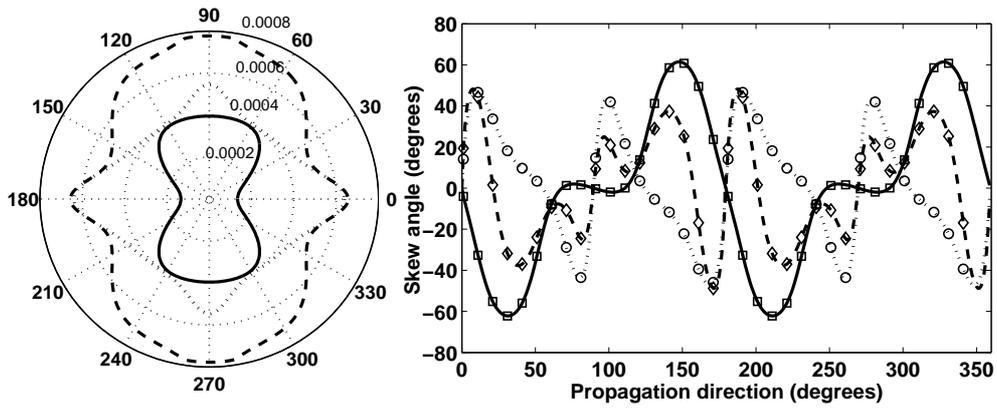}
  \caption{Left: Wave slowness curves for the P$_0$ (--), SH$_0$ ($\cdots$) and A$_0$ (-\,-) waves propagating in the orthotropic graphite-epoxy monolithic structure at 0.5MHz. Right: Corresponding energy skew angles computed through the presented approach for the P$_0$ (--), SH$_0$ ($\cdots$) and A$_0$ (-\,-) waves. Also presented the skew angles computed through a FD scheme as exhibited in \cite{wang2007group} for the P$_0$ ($\square$), SH$_0$ ($\circ$) and A$_0$ ($\diamond$) waves.}
\label{fr2_2.eps}
\end{figure}

In order to validate the accuracy of the above presented approach, an orthotropic graphite-epoxy monolithic structure is modelled through FEs and the characteristics of the acoustic waves propagating within the structure are computed in a broadband frequency range. The mechanical characteristics of the structure are given through the following elastic stiffness matrix
\begin{equation}
\mathbf{C}_0=10^9\left[ {\begin{array}{cccccc}
94 &7.4&8.2 &0&0 &0 \\
7.4 &13 &9.1 &0&0 &0 \\
8.2 &9.1 &34 &0&0 &0  \\
0&0 &0  &3.6&0 &0  \\
0&0 &0  &0&7.2 &0  \\
0&0 &0  &0&0 &4.2  \end{array} } \right] \textnormal{N/m$^2$} \nonumber
\end{equation}
\noindent
while the density of the structure is $\rho$=1600kg/m$^3$ and its thickness is $h$=1mm. The dimensions of the modelled periodic segment are $L_x$=$L_y$=10mm with a mesh comprising 10 elements in each direction.  The results on the slowness curves as well as on the energy skew angles are presented in Figs.\ref{fr1_2.eps} and \ref{fr2_2.eps} at frequencies of 0.1MHz and 0.5MHz respectively. The results are compared to a FD scheme \cite{wang2007group} in which the group velocity at a given wave propagation direction is determined as
\begin{equation}
\displaystyle\frac{\partial \omega_{w} }{\partial k_{w}}=\lim_{\omega_{w2} \to \omega_{w1}} \frac{\omega_{w2} - \omega_{w1} }{k_{w2} - k_{w1}}
\end{equation}
\noindent
while a similar finite central difference scheme is employed for calculating the angular dependence of the frequency at which a certain wavenumber occurs
\begin{equation}
\displaystyle\frac{\partial \omega_{w} }{\partial \theta}=\lim_{\delta \theta \to 0} \frac{\omega_{w}(k)\mid_{\theta_1+\delta \theta/2}-\omega_{w}(k)\mid_{\theta_1-\delta \theta/2} }{\delta \theta}
\end{equation}
\noindent
Acceptable values for $\omega_{w2}$ and $\delta \theta$ should be derived through a relative error convergence study with $\omega_{w2}-\omega_{w1}$ and $\delta \theta$ gradually diminishing until the relative difference in the acquired results is inferior to a defined tolerance.

It is stressed that the scheme proposed in this work is able to compute the wavenumber angular sensitivity (and subsequently the energy skew angle) by determining and post-processing a single solution of the system. This overcomes the two primary drawbacks of FD approaches; the first being that FD schemes require multiple solutions of the system for computing each gradient (more accurate FD schemes such as centered second and higher order ones ask for three or five solutions for computing just a single gradient). The second drawback that is overcome by the presented approach is that the variable perturbation for a FD scheme should be determined through a solution convergence study which also requires multiple solutions of the system under investigation. When it comes to large industrial models comprising an important number of elements, FD schemes are therefore expected to be computationally cumbersome. In that case the approach presented herein is deemed more appropriate, providing simultaneous efficiency and accuracy advantages.

The results in Figs.\ref{fr1_2.eps} and \ref{fr2_2.eps} unveil the intense angular, frequency and wave-type dependence of the slowness curves for the three propagating elastic waves. The SH$_0$ wave velocity appears to converge towards the A$_0$ phase velocities in the 'stiffer' direction of the structure. The intense variation of the energy skewing effect is also demonstrated in the same figures with the maximum skew angle being greater than $55^o$ for all wave types. Due to the symmetry of the slowness curves all skew angles are $\psi$=0 at $\theta=0^o/180^o$ as well as at $\theta=90^o/270^o$. It is observed that the skew angle for the pressure wave is almost insensitive to frequency changes, while the skewing effect for the A$_0$ wave is much more intense around $\theta=0^o/180^o$ for higher frequencies. Moreover, an excellent correlation is observed between the exhibited computational scheme and the FD scheme.

\begin{figure}
  \centering
\includegraphics[clip,width=7cm]{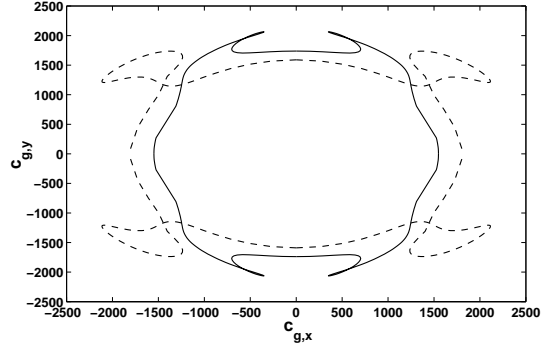}
  \caption{Group velocity curves for the A$_0$ (--) and the SH$_0$ (-\,-) waves propagating in the orthotropic graphite-epoxy monolithic structure, visualizing the appearance of caustics at 0.1MHz.}
\label{caustics1.eps}
\end{figure}

It should be noted that through the knowledge of the amplitude and actual direction of $c_g$ it is also straightforward to determine and visualize the appearance of caustics \cite{spadoni2009phononic} in the group velocity diagrams. An example of this wave behaviour is exhibited in Fig.\ref{caustics1.eps} for the A$_0$ and SH$_0$ propagating guided waves.

\section{Conclusions}
\label{sec:Conclusions}

The principal outcomes of the work are summarized as follows:

(i) A generic expression quantifying the wave energy skew angle as a function of the mechanical characteristics of a non-isotropic structure has been derived in this study. The approach does not involve any FD procedure and is therefore efficient and not prone to the associated numerical errors.

(ii) A FE-based approach for efficiently computing the angular sensitivity of the wave slowness for each wave type, direction and frequency was employed. The considered structure can be of arbitrary layering and material characteristics as an FE modelling approach is adopted. By employing periodic structure theory the associated computational effort is radically reduced.

(iii) An intense frequency dependence of the energy skew angle was observed for the A$_0$ waves travelling in an orthotropic graphite-epoxy monolithic structure. Angular and wave-type dependence was observed for the entirety of propagating waves with the skew angle being as pronounced as $65^o$ in some cases. It was also shown that the presented approach can successfully determine and visualize the appearance of caustics in the group velocity curves.





\bibliographystyle{model1-num-names}
\bibliography{wave_skewing}

\begin{thebibliography}{29}
\expandafter\ifx\csname natexlab\endcsname\relax\def\natexlab#1{#1}\fi
\providecommand{\bibinfo}[2]{#2}
\ifx\xfnm\relax \def\xfnm[#1]{\unskip,\space#1}\fi
\bibitem[{Yan et~al.(2010)Yan, Royer, and Rose}]{yan2010ultrasonic}
\bibinfo{author}{F.~Yan}, \bibinfo{author}{R.~L. Royer}, \bibinfo{author}{J.~L.
  Rose},
\newblock \bibinfo{title}{Ultrasonic guided wave imaging techniques in
  structural health monitoring},
\newblock \bibinfo{journal}{Journal of intelligent material Systems and
  Structures} \bibinfo{volume}{21} (\bibinfo{year}{2010})
  \bibinfo{pages}{377--384}.
\bibitem[{Kersemans et~al.(2014)Kersemans, Van~Paepegem, Van Den~Abeele, Pyl,
  Zastavnik, Sol, and Degrieck}]{kersemans2014pitfalls}
\bibinfo{author}{M.~Kersemans}, \bibinfo{author}{W.~Van~Paepegem},
  \bibinfo{author}{K.~Van Den~Abeele}, \bibinfo{author}{L.~Pyl},
  \bibinfo{author}{F.~Zastavnik}, \bibinfo{author}{H.~Sol},
  \bibinfo{author}{J.~Degrieck},
\newblock \bibinfo{title}{Pitfalls in the experimental recording of ultrasonic
  (backscatter) polar scans for material characterization},
\newblock \bibinfo{journal}{Ultrasonics} \bibinfo{volume}{54}
  (\bibinfo{year}{2014}) \bibinfo{pages}{1509--1521}.
\bibitem[{Lean and Chen(1979)}]{lean1979large}
\bibinfo{author}{E.~Lean}, \bibinfo{author}{W.~Chen},
\newblock \bibinfo{title}{Large-angle acoustic-beam steering in acoustically
  anisotropic crystal},
\newblock \bibinfo{journal}{Applied Physics Letters} \bibinfo{volume}{35}
  (\bibinfo{year}{1979}) \bibinfo{pages}{101--103}.
\bibitem[{Turnbull and Foster(1991)}]{turnbull1991beam}
\bibinfo{author}{D.~H. Turnbull}, \bibinfo{author}{F.~S. Foster},
\newblock \bibinfo{title}{Beam steering with pulsed two-dimensional transducer
  arrays},
\newblock \bibinfo{journal}{Ultrasonics, Ferroelectrics, and Frequency Control,
  IEEE Transactions on} \bibinfo{volume}{38} (\bibinfo{year}{1991})
  \bibinfo{pages}{320--333}.
\bibitem[{Smith et~al.(1991)Smith, Pavy~Jr, and Von~Ramm}]{smith1991high}
\bibinfo{author}{S.~W. Smith}, \bibinfo{author}{H.~G. Pavy~Jr},
  \bibinfo{author}{O.~T. Von~Ramm},
\newblock \bibinfo{title}{High-speed ultrasound volumetric imaging system. i.
  transducer design and beam steering},
\newblock \bibinfo{journal}{Ultrasonics, Ferroelectrics, and Frequency Control,
  IEEE Transactions on} \bibinfo{volume}{38} (\bibinfo{year}{1991})
  \bibinfo{pages}{100--108}.
\bibitem[{Clay et~al.(1999)Clay, Wooh, Azar, and Wang}]{clay1999experimental}
\bibinfo{author}{A.~C. Clay}, \bibinfo{author}{S.-C. Wooh},
  \bibinfo{author}{L.~Azar}, \bibinfo{author}{J.-Y. Wang},
\newblock \bibinfo{title}{Experimental study of phased array beam steering
  characteristics},
\newblock \bibinfo{journal}{Journal of Nondestructive Evaluation}
  \bibinfo{volume}{18} (\bibinfo{year}{1999}) \bibinfo{pages}{59--71}.
\bibitem[{Wooh and Shi(1999)}]{wooh1999optimum}
\bibinfo{author}{S.-C. Wooh}, \bibinfo{author}{Y.~Shi},
\newblock \bibinfo{title}{Optimum beam steering of linear phased arrays},
\newblock \bibinfo{journal}{Wave motion} \bibinfo{volume}{29}
  (\bibinfo{year}{1999}) \bibinfo{pages}{245--265}.
\bibitem[{Salas and Cesnik(2009)}]{salas2009guided}
\bibinfo{author}{K.~Salas}, \bibinfo{author}{C.~Cesnik},
\newblock \bibinfo{title}{Guided wave structural health monitoring using clover
  transducers in composite materials},
\newblock \bibinfo{journal}{Smart Materials and Structures}
  \bibinfo{volume}{19} (\bibinfo{year}{2009}) \bibinfo{pages}{1--25}.
\bibitem[{Wu et~al.(2012)Wu, Wu, Huang, and Yang}]{wu2012fpga}
\bibinfo{author}{S.~Wu}, \bibinfo{author}{M.~Wu}, \bibinfo{author}{C.~Huang},
  \bibinfo{author}{J.~Yang},
\newblock \bibinfo{title}{Fpga-based implementation of steerable parametric
  loudspeaker using fractional delay filter},
\newblock \bibinfo{journal}{Applied Acoustics} \bibinfo{volume}{73}
  (\bibinfo{year}{2012}) \bibinfo{pages}{1271--1281}.
\bibitem[{Ruzzene et~al.(2003)Ruzzene, Scarpa, and Soranna}]{ruzzene2003wave}
\bibinfo{author}{M.~Ruzzene}, \bibinfo{author}{F.~Scarpa},
  \bibinfo{author}{F.~Soranna},
\newblock \bibinfo{title}{Wave beaming effects in two-dimensional cellular
  structures},
\newblock \bibinfo{journal}{Smart materials and structures}
  \bibinfo{volume}{12} (\bibinfo{year}{2003}) \bibinfo{pages}{363--372}.
\bibitem[{Hussein et~al.(2008)Hussein, Leamy, and Ruzzene}]{hussein2008wave}
\bibinfo{author}{M.~I. Hussein}, \bibinfo{author}{M.~J. Leamy},
  \bibinfo{author}{M.~Ruzzene},
\newblock \bibinfo{title}{Wave beaming in nanostructured materials with
  engineered defects},
\newblock in: \bibinfo{booktitle}{ASME 2008 International Mechanical
  Engineering Congress and Exposition}, \bibinfo{organization}{American Society
  of Mechanical Engineers}, pp. \bibinfo{pages}{1011--1018}.
\bibitem[{Casadei and Rimoli(2013)}]{casadei2013anisotropy}
\bibinfo{author}{F.~Casadei}, \bibinfo{author}{J.~Rimoli},
\newblock \bibinfo{title}{Anisotropy-induced broadband stress wave steering in
  periodic lattices},
\newblock \bibinfo{journal}{International Journal of Solids and Structures}
  \bibinfo{volume}{50} (\bibinfo{year}{2013}) \bibinfo{pages}{1402--1414}.
\bibitem[{Newberry and Thompson(1989)}]{newberry1989paraxial}
\bibinfo{author}{B.~P. Newberry}, \bibinfo{author}{R.~B. Thompson},
\newblock \bibinfo{title}{A paraxial theory for the propagation of ultrasonic
  beams in anisotropic solids},
\newblock \bibinfo{journal}{The Journal of The Acoustical Society of America}
  \bibinfo{volume}{85} (\bibinfo{year}{1989}) \bibinfo{pages}{2290--2300}.
\bibitem[{Rose(2004)}]{rose2004ultrasonic}
\bibinfo{author}{J.~L. Rose}, \bibinfo{title}{Ultrasonic waves in solid media},
  \bibinfo{publisher}{Cambridge university press}, \bibinfo{year}{2004}.
\bibitem[{Wang and Yuan(2007)}]{wang2007group}
\bibinfo{author}{L.~Wang}, \bibinfo{author}{F.~Yuan},
\newblock \bibinfo{title}{Group velocity and characteristic wave curves of lamb
  waves in composites: Modeling and experiments},
\newblock \bibinfo{journal}{Composites Science and Technology}
  \bibinfo{volume}{67} (\bibinfo{year}{2007}) \bibinfo{pages}{1370--1384}.
\bibitem[{Carcione(2007)}]{carcione2007wave}
\bibinfo{author}{J.~M. Carcione}, \bibinfo{title}{Wave fields in real media:
  Wave propagation in anisotropic, anelastic, porous and electromagnetic
  media}, volume~\bibinfo{volume}{38}, \bibinfo{publisher}{Elsevier},
  \bibinfo{year}{2007}.
\bibitem[{Nelson(1976)}]{nelson1976simplified}
\bibinfo{author}{R.~B. Nelson},
\newblock \bibinfo{title}{Simplified calculation of eigenvector derivatives},
\newblock \bibinfo{journal}{AIAA journal} \bibinfo{volume}{14}
  (\bibinfo{year}{1976}) \bibinfo{pages}{1201--1205}.
\bibitem[{Adhikari and Friswell(2001)}]{adhikari2001eigenderivative}
\bibinfo{author}{S.~Adhikari}, \bibinfo{author}{M.~I. Friswell},
\newblock \bibinfo{title}{Eigenderivative analysis of asymmetric
  non-conservative systems},
\newblock \bibinfo{journal}{International Journal for Numerical Methods in
  Engineering} \bibinfo{volume}{51} (\bibinfo{year}{2001})
  \bibinfo{pages}{709--733}.
\bibitem[{Juang et~al.(1989)Juang, Ghaemmaghami, and Lim}]{juang1989eigenvalue}
\bibinfo{author}{J.-N. Juang}, \bibinfo{author}{P.~Ghaemmaghami},
  \bibinfo{author}{K.~B. Lim},
\newblock \bibinfo{title}{Eigenvalue and eigenvector derivatives of a
  nondefective matrix},
\newblock \bibinfo{journal}{Journal of Guidance, Control, and Dynamics}
  \bibinfo{volume}{12} (\bibinfo{year}{1989}) \bibinfo{pages}{480--486}.
\bibitem[{Friswell(1996)}]{friswell1996derivatives}
\bibinfo{author}{M.~Friswell},
\newblock \bibinfo{title}{The derivatives of repeated eigenvalues and their
  associated eigenvectors},
\newblock \bibinfo{journal}{Journal of vibration and acoustics}
  \bibinfo{volume}{118} (\bibinfo{year}{1996}) \bibinfo{pages}{390--397}.
\bibitem[{Finnveden(2004)}]{finnveden2004evaluation}
\bibinfo{author}{S.~Finnveden},
\newblock \bibinfo{title}{Evaluation of modal density and group velocity by a
  finite element method},
\newblock \bibinfo{journal}{Journal of Sound and Vibration}
  \bibinfo{volume}{273} (\bibinfo{year}{2004}) \bibinfo{pages}{51--75}.
\bibitem[{Cotoni et~al.(2008)Cotoni, Langley, and Shorter}]{Cotoni}
\bibinfo{author}{V.~Cotoni}, \bibinfo{author}{R.~S. Langley},
  \bibinfo{author}{P.~J. Shorter},
\newblock \bibinfo{title}{A statistical energy analysis subsystem formulation
  using finite element and periodic structure theory},
\newblock \bibinfo{journal}{Journal of Sound and Vibration}
  \bibinfo{volume}{318} (\bibinfo{year}{2008}) \bibinfo{pages}{1077--1108}.
\bibitem[{Ichchou et~al.(2007)Ichchou, Akrout, and Mencik}]{ichchou2007guided}
\bibinfo{author}{M.~Ichchou}, \bibinfo{author}{S.~Akrout},
  \bibinfo{author}{J.-M. Mencik},
\newblock \bibinfo{title}{Guided waves group and energy velocities via finite
  elements},
\newblock \bibinfo{journal}{Journal of Sound and Vibration}
  \bibinfo{volume}{305} (\bibinfo{year}{2007}) \bibinfo{pages}{931--944}.
\bibitem[{Chronopoulos(2015)}]{chronopoulos2015design}
\bibinfo{author}{D.~Chronopoulos},
\newblock \bibinfo{title}{Design optimization of composite structures operating
  in acoustic environments},
\newblock \bibinfo{journal}{Journal of Sound and Vibration}
  \bibinfo{volume}{355} (\bibinfo{year}{2015}) \bibinfo{pages}{322--344}.
\bibitem[{Chronopoulos et~al.(2014)Chronopoulos, Ichchou, Troclet, and
  Bareille}]{chronopoulos2014predicting}
\bibinfo{author}{D.~Chronopoulos}, \bibinfo{author}{M.~Ichchou},
  \bibinfo{author}{B.~Troclet}, \bibinfo{author}{O.~Bareille},
\newblock \bibinfo{title}{Predicting the broadband response of a layered
  cone-cylinder-cone shell},
\newblock \bibinfo{journal}{Composite Structures} \bibinfo{volume}{107}
  (\bibinfo{year}{2014}) \bibinfo{pages}{149--159}.
\bibitem[{Chronopoulos et~al.(2017)Chronopoulos, Collet, and
  Ichchou}]{chronopoulos2017wave}
\bibinfo{author}{D.~Chronopoulos}, \bibinfo{author}{M.~Collet},
  \bibinfo{author}{M.~Ichchou},
\newblock \bibinfo{title}{Wave sensitivity analysis for periodic and
  arbitrarily complex composite structures},
\newblock \bibinfo{journal}{Engineering Computations}  (\bibinfo{year}{2017})
  \bibinfo{pages}{00--}.
\bibitem[{Jones(1975)}]{jones1975mechanics}
\bibinfo{author}{R.~M. Jones}, \bibinfo{title}{Mechanics of composite
  materials}, volume \bibinfo{volume}{193}, \bibinfo{publisher}{Scripta Book
  Company Washington, DC}, \bibinfo{year}{1975}.
\bibitem[{Spadoni et~al.(2009)Spadoni, Ruzzene, Gonella, and
  Scarpa}]{spadoni2009phononic}
\bibinfo{author}{A.~Spadoni}, \bibinfo{author}{M.~Ruzzene},
  \bibinfo{author}{S.~Gonella}, \bibinfo{author}{F.~Scarpa},
\newblock \bibinfo{title}{Phononic properties of hexagonal chiral lattices},
\newblock \bibinfo{journal}{Wave motion} \bibinfo{volume}{46}
  (\bibinfo{year}{2009}) \bibinfo{pages}{435--450}.
\bibitem[{Langley(1993)}]{Langley1993377}
\bibinfo{author}{R.~Langley},
\newblock \bibinfo{title}{A note on the force boundary conditions for
  two-dimensional periodic structures with corner freedoms},
\newblock \bibinfo{journal}{Journal of Sound and Vibration}
  \bibinfo{volume}{167} (\bibinfo{year}{1993}) \bibinfo{pages}{377--381}.

\end{thebibliography}

\appendix
\renewcommand\thefigure{\arabic{figure}}
\section{Determining the angular sensitivity of the propagating wave characteristics through a finite element scheme}
\label{sec:vityanal}
\subsection{Computation of propagating wave properties through a finite element approach}
The wave propagation analysis scheme presented below has been first exhibited in \cite{Langley1993377}. The DoF set ${\bf q}$ (as well as the $\mathbb{M}$, $\mathbb{K}$ matrices) is reordered according to a predefined sequence such as:
\begin{equation}
\begin{array}{cc}
{\bf q}=\left\{ {\bf q_I \ \ q_B \ \ q_T \ \ q_L \ \ q_R \ \ q_{LB} \ \ q_{RB} \ \ q_{LT} \ \ q_{RT}} \right\}^{\top}
\end{array}
\end{equation}
\noindent
corresponding to the internal, the interface edge and the interface corner DoF (see Fig.\ref{fig:stif_panel.eps}). The free harmonic vibration equation of motion for the modelled segment is written as:
\begin{equation}
\begin{array}{cc}
[\bf{\mathbb{K}}-\omega^2\bf{\mathbb{M}}]\bf{q}=\bf{0}
\end{array}
\end{equation}
\noindent
The analysis then follows as in \cite{Cotoni} with the following relations being assumed for the displacement DoF under the passage of a time-harmonic wave:
\begin{equation}
\begin{array}{cc}
{\bf q_R}=$e$^{-i \varepsilon_x} {\bf q_L}, \ \ {\bf q_T}=$e$^{-i \varepsilon_y} {\bf q_B} \\
{\bf q_{RB}}=$e$^{-i \varepsilon_x} {\bf q_{LB}},  \ \ {\bf q_{LT}}=$e$^{-i \varepsilon_y} {\bf q_{LB}},  \ \ {\bf q_{RT}}=$e$^{-i \varepsilon_x -i \varepsilon_y} {\bf q_{LB}}
\end{array}
\label{eq:five}
\end{equation}
\noindent
with $\varepsilon_x$ and $\varepsilon_y$ the propagation constants in the $x$ and $y$ directions related to the phase difference between the sets of DoF. The wavenumbers $k_x$, $k_y$ are directly related to the propagation constants through the relation:
\begin{equation}
\begin{array}{cc}

\varepsilon_x=k_x L_x, \ \ \varepsilon_y=k_y L_y

\end{array}
\end{equation}
\noindent
Considering Eq.\ref{eq:five} in tensorial form gives:
\noindent
\begin{equation}
\begin{array}{cc}

{\bf q}=
\left[ {\begin{array}{cccc}
{\bf I} & {\bf 0} & {\bf 0} & {\bf 0} \\

{\bf 0} & {\bf I} & {\bf 0} & {\bf 0} \\
{\bf 0} & {\bf I} $e$^{-i \varepsilon_y} & {\bf 0} & {\bf 0} \\
{\bf 0} & {\bf 0} & {\bf I} & {\bf 0} \\
{\bf 0} & {\bf 0} & {\bf I}$e$^{-i \varepsilon_x} & {\bf 0} \\

{\bf 0} & {\bf 0} & {\bf 0} & {\bf I} \\
{\bf 0} & {\bf 0} & {\bf 0} & {\bf I}$e$^{-i \varepsilon_x} \\
{\bf 0} & {\bf 0} & {\bf 0} & {\bf I}$e$^{-i \varepsilon_y} \\
{\bf 0} & {\bf 0} & {\bf 0} & {\bf I}$e$^{-i \varepsilon_x -i \varepsilon_y} \\
\end{array} } \right]
\mathbf{x}={\bf R} \mathbf{x}

\end{array}
\end{equation}
\noindent
with $\mathbf{x}$ the reduced set of DoF: $\mathbf{x}=\left\{ {\bf q_I \ \ q_B \ \ q_L \ \ q_{LB} } \right\}^\top$. The equation of free harmonic vibration of the modelled segment can now be written as:

\begin{equation}
\begin{array}{cc}

[{\bf R}^*\mathbb{K}{\bf R}-\omega^2{\bf R}^*\mathbb{M}{\bf R}] \mathbf{x}=\bf{0}

\end{array}
\label{eq:eigen}
\end{equation}
\noindent
with $^*$ denoting the Hermitian transpose. The most practical procedure for extracting the wave propagation characteristics of the segment from Eq.\ref{eq:eigen} is injecting a set of assumed propagation constants $\varepsilon_x$, $\varepsilon_y$. The set of these constants can be chosen in relation to the direction of propagation towards which the wavenumbers are to be sought and according to the desired resolution of the wavenumber curves. Eq.\ref{eq:eigen} is then transformed into a standard eigenvalue problem and can be solved for the eigenvector $\mathbf{x}_w$ which describe the deformation of the segment under the passage of each wave type $w$ at an angular frequency equal to the square root of the corresponding eigenvalue $\lambda_w=\omega_w^2$. A complete description of each passing wave including its $x$ and $y$ directional wavenumbers and its wave shape for a certain frequency is therefore acquired. It is noted that the periodicity condition is defined modulo 2$\pi$, therefore solving Eq.\ref{eq:eigen} with a set of $\varepsilon_x$, $\varepsilon_y$ varying from 0 to 2$\pi$ will suffice for capturing the entirety of the structural waves. Further considerations on reducing the computational expense of the problem are discussed in \cite{Cotoni}.

\end{document}